\documentclass[12pt]{article}
\usepackage{amsmath,amsfonts}
\numberwithin{equation}{section}
\newcommand{\Cb}{{\mathbb C}}
\newcommand{\Cx}{{\mathbb C}^{\times}}
\newcommand{\Ncal}{{\cal N}}
\newcommand{\del}{\partial}

\newcommand{\deltam}[2]{\delta^{\rm mod\;#2}_{#1}}
\newcommand{\ch}{\hat{c}}

\newcommand{\Th}{\Theta}
\newcommand{\Zb}{{\mathbb Z}}
\newcommand{\e}[1]{\,{\bf e}\!\left[#1\right]}

\newcommand{\Rb}{{\mathbb R}}
\newcommand{\lcm}{{\rm lcm}}
\newcommand{\mt}{\underset{\widetilde{}}{\mu}}
\newcommand{\nt}{\underset{\widetilde{}}{\nu}}
\newcommand{\bt}[1]{\underset{\widetilde{}}{\beta_{#1}}}
\newcommand{\gt}{\underset{\widetilde{}}{\gamma_{0}}}

\newcommand{\taub}{\bar{\tau}}
\newcommand{\lambdab}{\bar{\lambda}}
\newcommand{\nn}{\nonumber}
\newcommand{\Zh}{\hat{Z}}
\newcommand{\Gh}{\hat{G}}
\newcommand{\gh}{\hat{g}}
\newcommand{\Tr}{{\rm Tr}}
\renewcommand{\atop}[2]{%
\genfrac{}{}{0pt}{}{#1}{#2}}
\begin{document}
\baselineskip 3.3ex

%
%
\newcommand{\KUCPlogo}{\hbox{\lower 1.4ex\hbox{\Huge\boldmath $\cal K$}
\kern -1.15em {\sffamily \bfseries\large\ UCP}
\put(-27.5,-6){\tiny\it preprint}
}}
\newcommand{\preprintnumber}{KUCP-0162 \\
 {\tt hep-th/0007069}}

\thispagestyle{empty}
\hbox{\raise 0.3ex\KUCPlogo\hspace{11cm}
\parbox[b]{4cm}{\preprintnumber}}
\noindent\hbox{\raise 1.5 ex \hbox{\rule{\textwidth}{0.3pt}}}

\vspace{20mm}\noindent
{\sffamily \bfseries \LARGE \boldmath
 Gepner-like Description
of a String Theory\\ on a 
Noncompact Singular Calabi-Yau Manifold
}

\vspace{0.5cm}\noindent
\rule{\textwidth}{1pt}

\vspace{2cm}

\noindent
{\sffamily \bfseries \large Satoshi Yamaguchi}

\vspace{5mm}
\noindent
\hspace{0.7cm} \parbox{140mm}{\it
Graduate School of Human and Environmental Studies,
Kyoto University, Yoshida-Nihonmatsu-cho,
Sakyo-ku, Kyoto 606-8501, Japan.

\vspace*{3mm}
E-mail: {\tt yamaguch@phys.h.kyoto-u.ac.jp}
}

\vspace{30mm}

\noindent
{\sc Abstract:\ \ }
We investigate a Gepner-like superstring model described by a combination
of multiple minimal models and an $\Ncal=2$ Liouville theory.
This model is thought to be equivalent to the superstring theory
on a singular noncompact Calabi-Yau manifold. We construct the modular
invariant partition function of this model, and confirm the validity of
an appropriate GSO projection.
We also calculate the elliptic genus and Witten index of
the model. We find that the elliptic genus is factorised into
a rather trivial factor and a non-trivial one, 
 and the non-trivial one has the information on
the positively curved base manifold of the cone.

%

\newpage
%
%

\section{Introduction}
The correspondence between a string theory on a K\"ahler manifold
 and an $\Ncal=2$ Landau-Ginzburg theory
is interesting and is very largely investigated
\cite{Gep87,Gep88,Vaf89,IV90}. But, the most results are
limited to the cases of compact Calabi-Yau manifolds.
Recently, it is conjectured that
in the case of a noncompact Calabi-Yau manifold, the associated
CFT consists of an $\Ncal=2$ Liouville theory and a
Landau-Ginzburg theory
\cite{GKP9907}.
They claimed that when the
Calabi-Yau $n$-fold $X$ is written as a hypersurface
 $F(z_1,\dots,z_{n+1})=0$ in $\Cb^{n+1}$
by a quasi-homogeneous polynomial $F$, 
then the string theory on $X$ is equivalent to the CFT on
\begin{eqnarray*}
 && \Rb_{\phi}\times S^1 \times LG(W=F).
\end{eqnarray*}
Here $\Rb_{\phi}$ is a real line parametrized by $\phi$
with linear dilaton background and
 $LG(W=F)$ is the IR theory of the Landau-Ginzburg model with
the superpotential $F$. In this case, the background charge $Q$ 
of $\Rb_{\phi}$ is
determined by a condition of the total central charge.
From the condition that $Q$ is a non-zero real number, we find
that the base manifold $X/\Cx$ should be curved positively.

The boson with linear dilaton background is strongly coupled
in the region $\phi\to -\infty$, so we should introduce
the Liouville potential or consider an $SL(2)/U(1)$ Kazama-Suzuki
model to avoid the strong coupling singularity
\cite{OV9511,GK9909,GK9911}. But we do not care about this point
in this paper.

In \cite{ABKS9808,GKP9907,GK9909,GK9911}, they also claim that
the string theory on this singular noncompact Calabi-Yau
manifold $X$ is holographic dual to the ``little string theory''.

In the case of a compact Calabi-Yau manifold, 
the string theory is ``solved'' in the description
of Gepner model in a special point of the moduli space.
We want to describe also the string theory on the noncompact 
Calabi-Yau manifold $X$ by the Gepner-like solvable model.
If we can do it successfully, it will be possible to analyze
more deeply
a noncompact Calabi-Yau manifold and the little string theory.

In \cite{ES0002}, they treat the string theories with ADE simple
singularities.
 They construct the modular invariant
 partition functions, and show the consistency of these string theories.

In this paper, we consider more general cases, in which the
Landau-Ginzburg part is described by a direct product of
a number of minimal models.
A typical example of ours is the Calabi-Yau $n$-fold $X$
described in the form
\begin{eqnarray*}
 &&z_1^{N_1}+z_2^{N_2}+\dots+z_{n+1}^{N_{n+1}}=0, \mbox{ in }\Cb^{n+1}.
\end{eqnarray*}
We construct the modular invariant partition functions and show the 
string theory actually exists consistently in these cases. 
We also calculate
the elliptic genus, and find that it is factorised into
 two parts --- a rather trivial one
and a rather non-trivial one.
We analyze the non-trivial one in detail,
and find that it has the information 
on the cohomology of the positively curved 
base manifold $X/\Cx$ except the elements
generated by cup products of a K\"ahler form.

The organization of this paper is as follows. In the next section,
we explain the setup and review the correspondence between
a noncompact Calabi-Yau manifold and an $\Ncal=2$ Liouville theory
 $\times$ Landau-Ginzburg theory. In section \ref{Sec3}, we
construct the modular invariant partition function.
In section \ref{Sec4}, we calculate the elliptic genus and compare it with
the geometric property of the associated
 Calabi-Yau manifold $X$. In the last
section, we summarize the results and discuss the problems and prospects.
In Appendix A. we collect some useful equations of theta functions and
characters that we use in this paper.

\section{The string theory on a 
noncompact singular Calabi-Yau manifold}\label{Sec2}

We consider the string compactification to a noncompact, singular
Calabi-Yau $n$-fold $X$. The total target space is
expressed by a direct product of a $d$ dimensional flat spacetime
and the manifold $X$
\begin{eqnarray}
 && \Rb^{d-1,1}\times X. \label{TargetManifold}
\end{eqnarray}
Here, $n$ is related to $d$ by the constraint on the total
 dimension $2n+d=10$.

For simplicity, we concentrate the case that the noncompact
 singular Calabi-Yau manifold $X$ is realized as the 
hypersurface in $\Cb^{n+1}$ determined by the algebraic equation
with a quasi homogeneous polynomial $F$
\begin{eqnarray*}
 F(z_1,\dots,z_{n+1})=0.
\end{eqnarray*}
By the term ``quasi-homogeneous'',
we mean that the polynomial $F$ satisfies
\begin{eqnarray}
 F(\lambda^{r_1}z_1,\dots,\lambda^{r_{n+1}}z_{n+1})=
\lambda F(z_1,\dots,z_{n+1}),\label{QuasiHomogeneous}
\end{eqnarray}
for some exponents $\{r_j\}$ and for an arbitrary $\lambda\in \Cx$.

This manifold $X$ is singular at $(z_1,\dots,z_{n+1})=(0,\dots,0)$.
If we consider the manifold $(X-{(0,\dots,0)})/\Cx$,
where the action of $\Cx$ is $(z_1,\dots,z_{n+1})\to
(\lambda^{r_1}z_1,\dots,\lambda^{r_{n+1}}z_{n+1})$ with the exponents
$\{r_j\}$ of (\ref{QuasiHomogeneous}), then we get a compact manifold.
We denote this compact manifold simply as $X/\Cx$ and call it
``the base manifold of $X$''.

It is conjectured in \cite{GKP9907} that the string theory 
on the space (\ref{TargetManifold}) is equivalent
to the theory including flat spacetime $\Rb^{d-1,1}$, a line with
the linear dilation background $\Rb_{\phi}$, $S^1$, and the Landau-Ginzburg
theory with a superpotential $W=F$;
\begin{eqnarray*}
  \Rb^{d-1,1}\times \Rb_{\phi} \times S^1 \times LG(W=F).
\end{eqnarray*} 

The part $(\Rb_{\phi} \times S^1)$ has a world sheet $\Ncal=2$
superconformal symmetry. 
Let $\phi$ be the parameter of $\Rb_{\phi}$ , $Y$ be the parameter of
$S^1$, and $\psi^{+},\psi^{-}$ be the fermionic part 
of $(\Rb_{\phi} \times S^1)$.
The $\Ncal=2$ superconformal currents are written in terms of
 the above fields
\begin{eqnarray}
 && T=-\frac12(\del Y)^2-\frac12(\del\phi)^2-\frac Q2 \del^2 \phi
-\frac12(\psi^{+}\del\psi^{-}-\del \psi^{+}\psi^{-}),\nn\\
 &&G^{\pm}=-\frac{1}{\sqrt 2}\psi^{\pm}(i\del Y\pm\del\phi)\mp
\frac{Q}{\sqrt2}\del\psi^{\pm},\nn\\
 && J=\psi^{+}\psi^{-}- Qi\del Y. \label{LiouvilleSCA}
\end{eqnarray}
The associated central charge of this algebra is $\ch(=c/3)=1+Q^2$.

In this paper, we consider the case in which the Landau-Ginzburg
theory with superpotential $W=F$ can be described by a direct product of
 $\Ncal=2$ minimal models.
Let $M_{G,N}$ be the minimal model
corresponding to simply laced Lie algebra 
$G=A,D,E$ with dual Coxeter number $N$.
We consider the theory in the following;
\begin{eqnarray*}
 && \Rb^{d-1,1}\times \Rb_{\phi} \times S^1 \times M_{G_1,N_1}\times
\dots\times M_{G_R,N_R},
\end{eqnarray*}
where $R$ is the number of the minimal models. The cases with $R=1$
are treated in \cite{ES0002} and $R=0$ in \cite{Miz0003,Kut9110}

A typical example is the case that all $G_j$ are $A$ type. In this example,
the quasi-homogeneous polynomial is written as
\begin{eqnarray*}
 && F(z)=z_1^{N_1}+\dots+z_{R}^{N_R}+z_{R+1}^{2}+\dots+z_{n+1}^2.
\end{eqnarray*}

The background charge of $\Rb_{\phi}$ is determined by the criticality
condition. To cancel the conformal
anomaly, the total central charge is to be $0$. The central charge of the
ghost sector is $-15$, so the total central charge of the 
matter sector is to be $15$. 
The central charge of the flat spacetime is $3/2$ for each pair
of a boson and a fermion, 
and that of the $\Rb_{\phi}\times S^1$ is $3+3Q^2$ as
mentioned above, and that of a minimal model $M_{G,N}$ is
$\frac{3(N-2)}{N}$. Therefore, the criticality
condition leads to the equation
\begin{eqnarray*}
 && \frac{3d}{2}+3+3Q^2+\sum_{j=1}^{R}\frac{3(N_j-2)}{N_j}=15.
\end{eqnarray*}
From this criticality condition, we obtain the value of $Q^2$ as
\begin{eqnarray}
 && Q^2=4-\frac d2 -\sum_j\frac{(N_j-2)}{N_j}.\label{CritCond1}
\end{eqnarray}
By the condition $Q^2>0$ for a real number $Q$,
the right-hand side should be positive;
\begin{eqnarray}
 4-\frac d2 -\sum_j\frac{(N_j-2)}{N_j}>0. \label{FDCondition}
\end{eqnarray}
It is equivalent to
a condition that the singularity is in finite distance in
the moduli space of deformation of singular Calabi-Yau
manifold $X$ \cite{GKP9907,GVW9906}. 
In view of the base manifold $X/\Cx$,
the finite distance condition is equivalent to that $X/\Cx$
is positively curved.

\section{Modular invariant partition function}\label{Sec3}

Now, let us construct the modular invariant partition function.
We take the light-cone gauge, then the associated
 CFT to consider is
\begin{eqnarray*}
 && \Rb^{d-2}\times \Rb_{\phi} \times S^1 \times M_{G_1,N_1}\times
\dots\times M_{G_R,N_R}.
\end{eqnarray*}

The toroidal partition function can be separated into 2 parts:
the one $Z_{GSO}$ concerning to the GSO projection and the
other $Z_0$ not concerning to it. We construct
 the total partition function $Z$ as
\begin{eqnarray*}
 Z=\int \frac{d^2\tau}{\tau_2^2}Z_0(\tau,\taub)Z_{GSO}(\tau,\taub),
\end{eqnarray*}
where $\tau=\tau_1+i\tau_2$ is the moduli parameter of the torus,
and $d^2\tau/\tau_2^2$ is the modular invariant measure.

First, we study the rather easy part $Z_0$, then we investigate
the rather complicated part $Z_{GSO}$.

\subsection{GSO independent part of the partition function}

In this subsection, we discuss the $Z_0$ : the GSO independent part.
It is completely the same as that in \cite{ES0002}.

The $Z_0$ includes the contribution from
the flat spacetime bosonic coordinates $X^I,\ (I=2,\dots,d-1)$ 
and the linear dilation $\phi$.

The partition function of each flat spacetime
boson is represented
by the Dedekind eta function $\eta(\tau)$ as
\begin{eqnarray*}
 &&\frac{1}{\sqrt{\tau_2}|\eta(\tau)|^2}.
\end{eqnarray*}

The partition function of $\phi$ is defined as
$Z_L=\Tr q^{L_0-c_L/24}\bar q^{\bar L_0-c_L/24},
\ (q=\exp(2\pi i\tau),c_L=1+3Q^2)$ in the canonical formalism.
Here the trace is taken over delta function normalizable
primary fields
\begin{eqnarray}
 \exp(ip\phi), \quad p=-\frac{iQ}{2}+\ell, \quad \ell\in \Rb,
\label{PrincipleSeries}
\end{eqnarray}
and their excitations by oscillators. 
Then we obtain $Z_L$ as
\begin{eqnarray*}
 Z_{L}&=&\frac{1}{|\prod_{n=1}(1-q^n)|^2}\int dp
\exp\left[-4\pi\tau_2\left(\frac12 p^2+\frac i2 pQ-\frac{1+3Q^2}{24}\right)
\right]\\
&=& \frac{1}{\sqrt{\tau_2}|\eta(\tau)|^2},
\end{eqnarray*}
where the region of the integral of $p$ is as 
(\ref{PrincipleSeries}).
As a result, the partition function of $\phi$ is the same as that
of an ordinary boson. So we obtain $Z_0$ as the
partition function of effectively $(d-1)$ free bosons;
\begin{eqnarray*}
 && Z_0=\left(\frac{1}{\sqrt{\tau_2}|\eta(\tau)|^2}\right)^{d-1}.
\end{eqnarray*}

The primary fields of (\ref{PrincipleSeries})
correspond to ``Principal continuous series'' in terms 
of the representation
of $SL(2)$. To include the other sectors is an interesting problem
and is postponed as a future work.

\subsection{GSO dependent part of the partition function
: $d=2,6$ cases}

Now, let us proceed to the GSO dependent part $Z_{GSO}$. 
In this subsection, we treat $d=2,6$ cases.

This part includes $(d-2)$ flat 
spacetime fermions $\psi^{I},\ (I=2,\dots,d-1)$ 
, two free fermions $\psi^{\phi},\psi^{Y}$
associated to $\Rb_{\phi}\times S^1$, minimal models
$M_{G_1,N_1},\dots,M_{G_R,N_R}$ , and an $S^1$ boson $Y$. 
We combine the 
$d$ free fermions $\psi^{I},\ (I=2,\dots,d-1) ,\psi^{\phi}$ and 
$\psi^{Y}$
and construct the affine Lie algebra $\widehat{SO(d)}_{1}$. 
The Verma module of $\widehat{SO(d)}_{1}$
is characterized by an integer $s_0=0,1,2,3$ , which labels
the representations of $SO(d)$, that is scalar, spinor, vector, and cospinor,
respectively.

Let us turn to Verma modules of $\Ncal=2$ minimal models.
The Verma module of an $\Ncal=2$ minimal model
is specified with three indices $(\ell,m,s)$, which satisfy
the following conditions\cite{Gep88}
\begin{eqnarray}
 && \ell=0,\dots,N-2,\nn\\
 && m=0,\dots,2N-1,\nn\\
 && s=0,1,2,3,\nn\\
 && \ell+m+s\equiv 0 \mod 2. \label{EvenCondition}
\end{eqnarray}
We denote  $\chi^{\ell,s}_m(\tau,z)$ as the character
of the Verma module labeled by the set $(\ell,m,s)$.
Some properties of this character is collected in Appendix A.

The Verma module of the whole GSO dependent parts is specified by
the index $s_0$ of $\widehat{SO(d)}_{1}$ representation,
the indices $(\ell_j,m_j,s_j),\ (j=1,\dots,R)$ of the minimal models,
and the $S^1$ momentum $p$.
We combine these indices except $p$ into two vectors $\lambda,\mu$.
\begin{eqnarray*}
 && \lambda:=(\ell_1,\dots,\ell_R), \\
 && \mu:=(s_0;s_1,\dots,s_R;m_1,\dots,m_R).
\end{eqnarray*}
We shall introduce the inner product between $\mu$ and $\mu'$
as in \cite{Gep88}
\begin{eqnarray*}
 \mu\bullet\mu':=-\frac d2 \frac{s_0s_0'}{4}-\sum_{j=1}^{R} \frac{s_j s_j'}{4}
+\sum_{j=1}^{R}\frac{m_jm_j'}{2N_j}.\\
\end{eqnarray*}
Also it is convenient to 
introduce special vectors $\beta_0,\beta_j\;(j=1,\dots,R)$
\begin{eqnarray*}
&& \beta_0:=(1;1,\dots,1;1,\dots,1),\\
&& \beta_j:=(2;0,\dots,0,\underset{\underset{S_j}{\wedge}}{2}
,0,\dots0;0,\dots,0).\\
\end{eqnarray*}
Here the $\beta_0$ is the vector with all components $1$, and $\beta_j$ is 
the vector with $s_0$ and $s_j$ components $2$ and the others zero.

With these notations, the criticality condition (\ref{CritCond1}) can 
be written in a rather simple form as 
\begin{eqnarray}
 && Q^2=4(1+\beta_0\bullet\beta_0).  \label{CritCond2}
\end{eqnarray}
When we define an integer $K:=\lcm(2,N_j)$, $KQ^2$ is shown to be
an even integer because of the equations
 (\ref{CritCond1}) and (\ref{CritCond2}). 
Therefore, it is convenient to define an integer $J$ by the equation
\begin{eqnarray}
 && J :=2K(1+\beta_0\bullet\beta_0)\quad(=K Q^2/2).\label{DefJ}
\end{eqnarray}
In terms of $J$, the finite distance condition (\ref{FDCondition}) can
be expressed as $J>0$.

Now, let us consider the character of the Verma module $(\lambda,\mu,p)$,
\begin{eqnarray*}
 && \chi^{\lambda}_{\mu}(\tau) \frac{q^{\frac12 p^2}}{\eta(\tau)},
\end{eqnarray*}
where $\chi^{\lambda}_{\mu}(\tau)$ is the product of
 characters of the minimal models and
the $\widehat{SO(d)}_1$ character $\chi_{s_0}(\tau)$ of the
$s_0$ representation
\begin{eqnarray*}
 \chi^{\lambda}_{\mu}(\tau):=\chi_{s_0}(\tau)\chi_{m_1}^{\ell_1,s_1}(\tau)
\dots\chi_{m_R}^{\ell_R,s_R}(\tau).
\end{eqnarray*}

In this character, $\chi^{\lambda}_{\mu}(\tau)$ has good modular
properties, but $q^{\frac12 p^2}/\eta(\tau)$ has bad ones.
So, we will sum up the characters with respect to
 certain values of $p$ and
make the
modular properties good \cite{Miz0003}\cite{ES0002}.

Let us consider the GSO projection. 
By the GSO projection, 
we pick up the states with odd integral $U(1)$ charges of the $\Ncal=2$
superconformal symmetry.
The $U(1)$ charge of the states in the above Verma module is expressed as
\begin{eqnarray*}
 2\beta_0\bullet\mu+pQ=-\frac d2 \frac{s_0}{2}-\sum_j \frac{s_j}{2}
+\sum_j\frac{m_j}{N_j}+pQ. \\
\end{eqnarray*}
From the condition that this $U(1)$ charge must be an odd integer
 $(2u+1)$ with $u\in \Zb$,
the $S^1$ momentum $p$ is written as
\begin{eqnarray*}
 && p(u)=\frac 1Q\left(2u+1-2\beta_0\bullet\mu \right).
\end{eqnarray*}
If we sum up the characters for all $u\in \Zb$, we obtain the theta
function with a fractional level\cite{ES0002}, 
which does not have good modular properties.
So we perform the following trick.

Let us write $u=Jv+w$ with integers $v,w$ and sum up the characters
for $v\in \Zb$. Then the sum leads to the following theta function
\begin{eqnarray}
 && \sum_{v\in \Zb}
 q^{\frac12 p(u=Jv+w)^2}=\Th_{-2K\beta_0\bullet\mu+K(2w+1),KJ}(\tau).
\label{GSOCondition0}
\end{eqnarray}
Note that $-2K\beta_0\bullet\mu+K(2w+1)$ is an integer, and the above
theta function has good modular properties.

Now, including oscillator modes and other sectors, 
we can define the building blocks
$f^{\lambda}_{\mt}(\tau)$ by
\begin{eqnarray*}
 &&f^{\lambda}_{\mt}(\tau):=\chi^{\lambda}_{\mu}(\tau)
\Th_{M,KJ}(\tau)/\eta(\tau),\\
 && \mt:=(\mu,M), \ M\in \Zb_{2KJ}.
\end{eqnarray*}
We should use only the building
blocks $f^{\lambda}_{\mt}$ with the conditions
\begin{eqnarray}
 && M=-2K\beta_0\bullet\mu+K(2w+1) \mbox{ for }{}^{\exists} w \in \Zb,
\label{GSOCondition}\\
 && s_0\equiv s_1 \equiv \dots \equiv s_R \mod 2 .
\label{SpinCondition}
\end{eqnarray}
The condition (\ref{GSOCondition}) comes 
from the formula of (\ref{GSOCondition0}), and
the condition (\ref{SpinCondition}) implies that 
the boundary condition of the fermionic currents are the same in
all the sub-theories, i.e. they must be all in the NS sector, 
or all in the R sector.

The modular invariant partition function can be systematically
obtained by ``the beta method''\cite{Gep88}.

The inner product between of two vectors $\mt,\mt'$ is defined as
\begin{eqnarray*}
 \mt\bullet\mt':=\mu\bullet\mu'-\frac{MM'}{2KJ}.
\end{eqnarray*}
We also extend the vectors $\beta_0,\beta_j$ to $\bt{0},\bt{j}$
as
\begin{eqnarray*}
 && \bt{0}:=(\beta_0,-J),\\
 && \bt{j}:=(\beta_j,0),\\
\end{eqnarray*}
and evaluate the inner products of these $\bt{0},\bt{j}$ vectors
\begin{eqnarray}
 && \bt{0}\bullet\bt{0}=\beta_0\bullet\beta_0-\frac{J^2}{2KJ}=-1,\nn\\
 &&\bt{j}\bullet \bt{j}=\beta_j\bullet\beta_j=-\frac d2-1, \nn\\
 && \bt{j}\bullet \bt{0}=\frac12\left(-\frac d2-1\right). \label{BetaBeta}
\end{eqnarray}
Note that $\bt{0}\bullet\bt{0}$ is an odd integer,
$\bt{j}\bullet \bt{j}$ are even integers (Recall that we consider
the cases $d=2,6$), and $\bt{j}\bullet \bt{0}$ are integers.
Using these special vectors, the conditions 
(\ref{GSOCondition}) and (\ref{SpinCondition})
are written in a simple form
\begin{eqnarray}
 && 2\bt{0}\bullet\mt\in 2\Zb+1,\nn\\
 && \bt{j}\bullet\mt \in \Zb.\label{BetaConditions}
\end{eqnarray}
We call this condition ``the beta condition''.

Using these notations, and the modular transformation laws of
theta functions and $\Ncal=2$ characters written in
the Appendix A, we can calculate the modular transformation
laws of $f_{\mt}^{\lambda}$ as
\begin{eqnarray*}
 &&f_{\mt}^{\lambda}(\tau+1)=\e{\sum_j\frac{\ell_j(\ell_j+1)}{4N_j}-\frac12 \mt\bullet\mt-\frac1{24}\left(\sum_{j}\frac{N_j-2}{N_j}+\frac d2+1\right)}
f_{\mt}^{\lambda}(\tau),\\
 &&f_{\mt}^{\lambda}(-1/\tau)=
\sum_{\lambda',\mt'}^{\rm even}
A_{\lambda\lambda'}\left(\prod_j\frac{1}{\sqrt{8N_j}}\right)
\frac{1}{\sqrt{8KJ}}\e{\mt\bullet\mt'}f_{\mt'}^{\lambda'}(\tau),
\end{eqnarray*}
where  the sums of the $\lambda',\mu'$
are taken only for the range (\ref{EvenCondition}) and for
$M=0,\dots,2KJ-1$.
Especially we must impose the condition $\ell_j+m_j+s_j\equiv0\mod 2$
for each minimal model.
$A_{\lambda\lambda'}$ is the products of the $\widehat{SU(2)}_{N_j-2}$
 S matrices $A^{(N_j)}_{\ell_j\ell_j'}$;
\begin{eqnarray*}
  A_{\lambda\lambda'}
=\prod_j A^{(N_j)}_{\ell_j\ell_j'}
=\prod_j\sqrt{\frac{2}{N_j}}\sin \pi
\frac{(\ell_j+1)(\ell'_j+1)}{N_j},
\end{eqnarray*}
and we use
here and the rest of this paper the notation $\e{x}=\exp(2\pi i x)$.

Let us note that if a vector $\mt$ satisfies the beta condition
(\ref{BetaConditions}), the vector $\mt+b_0\bt{0}+\sum_jb_j\bt{j}$
for $b_0,b_j\in\Zb,\ (j=1,\dots,R)$
 also satisfies the beta condition by virtue of (\ref{BetaBeta}).
Using this fact, we define the function $F_{\mt}^{\lambda}$
for $(\lambda,\mt)$ which satisfies the beta conditions 
(\ref{BetaConditions}) as a sum of
 $f_{\mt+b_0\bt{0}+\sum_jb_j\bt{j}}^{\lambda}$'s as
\begin{eqnarray*}
 && F_{\mu}^{\lambda}(\tau)=\sum_{b_0,b_j}(-1)^{s_0+b_0}
f_{\mt+b_0\bt{0}+\sum_jb_j\bt{j}}^{\lambda}(\tau),
\end{eqnarray*}
where the sum is taken for $b_0\in \Zb_{2K}$ and $b_j\in \Zb_2$.
The sign $(-1)^{s_0+b_0}$ is $(-1)$ for the Ramond sector.
 
These functions have very good modular properties. Especially
by S transformation, the functions are mixed among those which
satisfy the beta condition;
\begin{eqnarray*}
 && F_{\mt}^{\lambda}(\tau+1)=
\e{\sum_j\frac{\ell_j(\ell_j+1)}{4N_j}
-\frac12 \mt\bullet\mt-\frac1{24}\left(\sum_{j}\frac{N_j-2}{N_j}+\frac d2+1\right)}F_{\mt}^{\lambda}(\tau),\\
 && F_{\mt}^{\lambda}(-1/\tau)=\sum_{\lambda',\mt'}^{\rm even,beta}
A_{\lambda\lambda'}\left(\prod_j\frac{1}{\sqrt{8N_j}}\right)
\frac{1}{\sqrt{8KJ}}\e{\mt\bullet\mt'}(-1)^{s_0+s_0'}
F_{\mt'}^{\lambda'}(\tau),
\end{eqnarray*}
where the sums of $\lambda',\mt'$ is taken for 
restricted subclass that satisfies
the conditions (\ref{EvenCondition}) 
and the beta condition (\ref{BetaConditions}).

With this function $F_{\mu}^{\lambda}$, 
we obtain the modular invariant $Z_{GSO}$
as
\begin{eqnarray*}
 &&Z_{GSO}(\tau,\taub) = \frac{1}{4^R\times 2K}\sum_{\lambda,\lambdab,\mt}^{\rm even,beta}
L_{\lambda\lambdab}F_{\mt}^{\lambda}(\tau)\bar F_{\mt}^{\lambdab}(\taub),\\
\end{eqnarray*}
where $L_{\lambda\lambdab}=\prod_jL_{\ell_j\bar\ell_j}^{(G_j,N_j-2)}$ is 
the product of $G_j=A,D,E$ type modular invariants
of $\widehat{SU(2)}_{N_j-2}$ \cite{CIZ87,CIZ87CMP,Kat87}.

We can check the modular invariance of the above partition function.

We expect from spacetime supersymmetry that the partition function
vanishes, or equivalently $F_{\mt}^{\lambda}(\tau)=0$. It is a
future work to check that it actually vanishes.

Here, we find a solution, but it may not be the only solution, and
there can be some variety of modular invariant partition function.
Actually, for $R=1$ case, there are many other solutions
associated with the other modular invariants of the theta system
\cite{ES0002}.

\subsection{GSO dependent part of the partition function : $d=4$ case}

In this subsection, we comment on the $d=4$ case.
To construct the modular invariant partition function in
the $d=4$ case, we combine the four fermions to
construct the affine currents
$\widehat{SO(2)}_1\times \widehat{SO(2)}_1$ and
label the the Verma module by indices $s_{-1}$ and $s_0$.
Then, the modular invariant partition function can be 
constructed in almost the same way as the $d=2,6$ cases.

First we define the vectors $\mt$'s and the inner product
between them as
\begin{eqnarray*}
 && \mt:=(s_{-1},s_0;s_1,\dots,s_R;m_1,\dots,m_R;M),\\
 && \mt\bullet\mt':=-\frac{s_{-1}s_{-1}'}{4}-\frac{s_{0}s_{0}'}{4}
-\sum_{j}\frac{s_{j}s_{j}'}{4}+\sum_{j}\frac{m_jm_j'}{2N_j}
-\frac{MM'}{2KJ}.
\end{eqnarray*}
It is convenient to introduce  special vectors $\bt{0}$, $\bt{j}$ and
$\bt{-1}$
\begin{eqnarray*}
 && \bt{0}=(1,1;1,\dots,1;1,\dots,1;M),\\
 && \bt{j}=(0,2;0,\dots,0,\underset{\underset{S_j}{\wedge}}{2}
,0,\dots,0;0,\dots,0;0),\ (j=1,\dots,R),\\
 && \bt{-1}=(2,2;0,\dots,0;0,\dots,0;0).\\
\end{eqnarray*}
Using these vectors, we can construct the building blocks
 $f^{\lambda}_{\mt}(\tau)$ as
\begin{eqnarray*}
 &&f^{\lambda}_{\mt}(\tau):=
\chi_{s_{-1}}(\tau)\chi_{s_{0}}(\tau)\chi_{m}^{\ell_1,s_1}(\tau)
\dots \chi_{m}^{\ell_R,s_R}(\tau)
\Th_{M,KJ}(\tau)/\eta(\tau),
\end{eqnarray*}
where $\chi_{s_{-1}}(\tau)$ and $\chi_{s_{0}}(\tau)$ 
are $\widehat{SO(2)}_1$
characters. Then the GSO conditions and the condition
of fermionic sectors are
\begin{eqnarray}
 && 2\bt{0}\bullet \mt\in 2 \Zb+1,\qquad
 \bt{j}\bullet \mt \in \Zb,\qquad
 \bt{-1}\bullet \mt  \in \Zb, \label{BetaCondition4}
\end{eqnarray}
and we can construct the modular invariant partition function
by the beta method in this case.
Next we introduce the function $F_{\mt}^{\lambda}(\tau)$ as
\begin{eqnarray*}
 && F_{\mt}^{\lambda}(\tau)=\sum_{b_0\in\Zb_{2K},b_j\in\Zb_2,b_{-1}\in\Zb_2}
(-1)^{b_0+s_0}
f_{\mt+b_0\bt{0}+\sum_j b_j\bt{j}+b_{-1}\bt{-1}}^{\lambda}(\tau),
\end{eqnarray*}
then we obtain
the GSO dependent part of the modular invariant partition function 
$Z_{GSO}$
\begin{eqnarray*}
 &&Z_{GSO}(\tau,\taub) = 
\frac{1}{4^R\times 4K}\sum_{\lambda,\lambdab,\mt}^{\rm even,beta}
L_{\lambda\lambdab}F_{\mt}^{\lambda}(\tau)\bar F_{\mt}^{\lambdab}(\taub).\\
\end{eqnarray*}
We can check the modular invariance of the above partition function.

\section{Elliptic genus}\label{Sec4}

In this section, we calculate the elliptic genus of the theory
\cite{KYY9306}. The definition of the elliptic genus is
\begin{eqnarray*}
 && Z(\tau,\taub,z):=\Tr_{RR}(-1)^{F}q^{L_0-c/24}\bar q^{\bar L_0-c/24}y^{J_0},
\end{eqnarray*}
where the trace is taken for the RR states, and $y=\exp(2\pi iz)$. This
elliptic genus has the following modular properties;
\begin{eqnarray}
 && Z(\tau+1,\taub+1,z)=Z(\tau,\taub,-z)=Z(\tau,\taub,z),\nn\\
 && Z(-1/\tau,-1/\taub,z/\tau)=\e{\frac{\ch}{2}\frac{z^2}{\tau}}
Z(\tau,\taub,z). \label{EllipticModular}
\end{eqnarray}

Here, we omit the contribution from the flat space time and
consider only the internal part describing
the Calabi-Yau $n$-fold $X$. We calculate its elliptic genus and
the Witten index, and discuss its geometrical interpretation.

Let us consider again ``the criticality condition'', in other
words ``the Calabi-Yau condition'' of $X$.
Here the total $\ch$ should be $n$ because we want the theory
that describes a Calabi-Yau $n$-fold. Therefore, the total $\ch$ of 
the $\Ncal=2$ Liouville and the minimal models should satisfy
the relations
\begin{eqnarray}
 n=\ch=1+Q^2+\sum_j\frac{N_j-2}{N_j}.\label{CalabiYauCondition}
\end{eqnarray}

We introduce the following
vectors with $R$ components $\{m_j\}$, and the inner product between them as
\begin{eqnarray*}
 && \nu:=(m_1,\dots,m_R),\\
 && \nu\bullet\nu':=\sum_j\frac{m_jm_j'}{2N_j}.\\
\end{eqnarray*}
We also introduce the special vector $\gamma_0$ with all components $2$
\begin{eqnarray*}
 &&\gamma_0:=(2,\dots,2).
\end{eqnarray*}
With these notations, the condition
(\ref{CalabiYauCondition}) becomes
\begin{eqnarray*}
 && Q^2=n-1-R+\gamma_0\bullet\gamma_0.
\end{eqnarray*}
Next we let $N:=\lcm(N_j)$, and define $J$ as
\begin{eqnarray}
 && \frac{2J}{N}:=Q^2. \label{DefJ2}
\end{eqnarray}
In this paper, we concentrate only the case that $(n-1-R)$ is even, then
in this case, $J$ is an integer. In terms of $J$, the finite distance
condition $Q^2>0$ can be written as $J>0$.

Because we want a Calabi-Yau CFT, we have to pick up only the states with
integral $U(1)$ charges. This condition is realized as the condition
\begin{eqnarray}
 && \gamma_0\bullet\nu+pQ \in \Zb. \label{GammaCondition0}
\end{eqnarray}
From this, $p$ can be written with an arbitrary
integer $u$
\begin{eqnarray*}
 && p=\frac 1Q \left(u-\gamma_0\bullet\nu\right).
\end{eqnarray*}
Following the same manner as in the previous section, 
we let $u=2Jv+w$ and sum up for $v\in\Zb$. It leads to the
theta function
\begin{eqnarray*}
 &&\sum_{v\in\Zb}q^{\frac12p^2}y^{pQ}=
\Th_{N(w-\gamma_0\bullet\nu),NJ}(\tau,2z/N).
\end{eqnarray*}
Note that $N(w-\gamma_0\bullet\nu)$ is an integer and
$\Th_{N(w-\gamma_0\bullet\nu),NJ}(\tau,2z/N)$ has good modular
properties.

Collecting these, we define the building blocks $g_{\nt}^{\lambda}$ as
\begin{eqnarray*}
 && g_{\nt}^{\lambda}(\tau,z):=\sum_{s_0,s_j=1,3}
\chi_{\mu}^{\lambda}(\tau,z)
\frac{\Th_{M,NJ}(\tau,2z/N)}{\eta(\tau)}
(-1)^{-\frac{s_0}{2}-\sum_j\frac{s_j}{2}
+\gamma_0\bullet\nu+\frac{M}{N}},
\end{eqnarray*}
where $\nt:=(\nu,M)$. In the sign
 $(-1)^{J_0}=(-1)^{-\frac{s_0}{2}-\sum_j\frac{s_j}{2}
+\gamma_0\bullet\nu+\frac{M}{N}}$ , the part
 $(-\frac{s_0}{2}-\sum_j\frac{s_j}{2})$ represents contributions 
of ordinary $U(1)$ charges
 from the indices $s_0,s_j$, and the rest
 $\gamma_0\bullet\nu+\frac{M}{N}=w=u-2Jv$ reflects contributions from
the indices $m_j$ and $S^1$ momentum.

Let us define the inner product between $\nt$ and $\nt'$ as
\begin{eqnarray*}
 && \nt\bullet\nt':=\nu\bullet\nu'-\frac{MM'}{2NJ},
\end{eqnarray*}
and the special vector
\begin{eqnarray*}
 && \gt=(\gamma_0,-2J).
\end{eqnarray*}
We also introduce the functions $I_{m}^{\ell}$ and $I_{\nu}^{\lambda}$
\begin{eqnarray*}
 && I_{m}^{\ell}(\tau,z):=\chi_{m}^{\ell,1}(\tau,z)
-\chi_{m}^{\ell,3}(\tau,z),\\
 && I_{\nu}^{\lambda}(\tau,z):=I_{m_1}^{\ell_1}(\tau,z)
\dots I_{m_R}^{\ell_R}(\tau,z).
\end{eqnarray*}
With these notations, the building blocks $g_{\nt}^{\lambda}$ can be
written as
\begin{eqnarray*}
 g_{\nt}^{\lambda}(\tau,z)
=\frac{\theta_1(\tau,z)}{\eta(\tau)}I_{\nu}^{\lambda}(\tau,z)
\frac{\Th_{M,NJ}(\tau,2z/N)}{\eta(\tau)}(-1)^{\gt\bullet\nt},
\end{eqnarray*}
where we omit the overall irrelevant phase.
The condition (\ref{GammaCondition0}) can be rewritten as
\begin{eqnarray}
 && \gt\bullet\nt\in \Zb,\label{GammaCondition}
\end{eqnarray}
and again we call this condition ``the beta condition''.

Now, we construct elliptic genus 
using the above building blocks $g_{\nt}^{\lambda}$ 
which satisfy the condition (\ref{GammaCondition}).

Note that if $\nt$ satisfies the beta condition, then $\nt+b_0\gt$ 
for $b_0\in \Zb$ also
satisfies the beta condition, because
\begin{eqnarray*}
 && \gt\bullet\gt=\gamma_0\bullet\gamma_0-\frac{2J}{N}=-(n-1-R),
\end{eqnarray*}
is an integer.
\footnote{
Actually, it is an even integer. Remember that we concentrate the case
in which $(n-1-R)$ is an even integer.}
Here we used the definition of $J$ (\ref{DefJ2}).
So let us define the new functions $G_{\nt}^{\ell}$
 as follows;
\begin{eqnarray*}
 && G_{\nt}^{\ell}(\tau,z)=\sum_{b_0\in \Zb_{N}}
g_{\nt+b_0\gt}^{\lambda}(\tau,z),
\end{eqnarray*}
where $\nt$ satisfies the beta condition (\ref{GammaCondition}).
Then, from the modular properties of $g_{\nt}^{\lambda}$
\begin{eqnarray*}
 && g_{\nt}^{\lambda}(\tau+1,z)=
\e{\sum_j\frac{\ell_j(\ell_j+1)}{4N_j}-\frac12 \nt\bullet\nt
+\frac{R+1}{8}-\frac1{24}\left(\sum_{j}\frac{N_j-2}{N_j}+2\right)}
g_{\nt}^{\lambda}(\tau,z),\\
 && g_{\nt}^{\lambda}(-1/\tau,z/\tau)=(-i)^R \e{\frac n2 \frac{z^2}{\tau}}
\\&&\hspace{2cm}\times
\sum_{\lambda',\nt'}^{\rm even}
A_{\lambda\lambda'}\frac{1}{\prod_j\sqrt{2N_j}}
\frac{1}{\sqrt{2NJ}}\e{\nt\bullet\nt'}(-1)^{\gt\bullet(\nt-\nt')}
g_{\nt'}^{\lambda'}(\tau,z),
\end{eqnarray*}
$G_{\nt}^{\ell}$ have very good modular properties;
\begin{eqnarray*}
 && G_{\nt}^{\lambda}(\tau+1,z)=
\e{\sum_j\frac{\ell_j(\ell_j+1)}{4N_j}-\frac12 \nt\bullet\nt
+\frac{R+1}{8}-\frac1{24}\left(\sum_{j}\frac{N_j-2}{N_j}+2\right)}
G_{\nt}^{\lambda}(\tau,z),\\
 && G_{\nt}^{\lambda}(-1/\tau,z/\tau)=(-i)^R \e{\frac n2 \frac{z^2}{\tau}}
\\&&\hspace{2cm}\times
\sum_{\lambda',\nt'}^{\rm even,beta}
A_{\lambda\lambda'}\frac{1}{\prod_j\sqrt{2N_j}}
\frac{1}{\sqrt{2NJ}}\e{\nt\bullet\nt'}(-1)^{\gt\bullet(\nt-\nt')}
G_{\nt'}^{\lambda'}(\tau,z).
\end{eqnarray*}
Using these functions, we obtain the elliptic genus in the following
form;
\begin{eqnarray*}
 && Z(\tau,\taub,z)=\frac1{2^{R}N}
\frac{1}{\sqrt{\tau_2}|\eta(\tau)|^2}
\sum_{\lambda,\lambdab,\nt}^{\rm even,beta}
L_{\lambda\lambdab}G_{\nt}^{\lambda}(\tau,z)\bar G_{\nt}^{\lambdab}(\taub,0).
\end{eqnarray*}
Here $L_{\lambda\lambdab}$ is the product of $\widehat{SU(2)}$ modular
invariants, and the factor $1/\sqrt{\tau_2}|\eta(\tau)|^2$ is
contribution of $\phi$. 
We can check that the above elliptic genus has the right modular
properties (\ref{EllipticModular}) with $\ch=n$.

Actually, this elliptic genus is $0$ because it has an overall factor
$\bar\theta_1(\taub,0)=0$.
\subsection{Hodge number and Witten index}
To get some nontrivial information from the above elliptic genus,
we factor out the trivial parts and define $\Zh$ by the equations
\begin{eqnarray*}
 && Z(\tau,\taub,z)=
\frac{\theta_1(\tau,z)\bar\theta_1(\tau,0)}{\sqrt{\tau_2}|\eta(\tau)|^6}
\Zh(\tau,\taub,z),\\
 && \Zh(\tau,\taub,z)=\frac1{2^{R}N}
\sum_{\lambda,\lambdab,\nt}^{\rm even,beta}
L_{\lambda\lambdab}\Gh_{\nt}^{\lambda}(\tau,z)
\bar{\Gh}_{\nt}^{\lambdab}(\taub,0),\\
&& \Gh_{\nt}^{\lambda}(\tau,z)=\sum_{b_0\in \Zb_{N}}
\gh_{\nt+b_0\gt}^{\lambda}(\tau,z),\\
&& \gh_{\nt}^{\lambda}(\tau,z)=I_{\nu}^{\lambda}(\tau,z)
\Th_{M,NJ}(\tau,2z/N)(-1)^{\gt\bullet\nt},
\end{eqnarray*}

Now, we take the limit
$\tau\to i \infty$ and consider the ground states.
In this limit, $\Th_{M,NJ}$ becomes
\begin{eqnarray*}
 \Th_{M,NJ}(i\infty,z)=\deltam{M}{NJ},
\end{eqnarray*}
so, the $\Gh$'s can be evaluated as
\begin{eqnarray*}
 &&\Gh_{\nt}^{\lambda}=
\begin{cases}
I_{\nu+\frac{M}{2J}\gamma_0}^{\lambda}(i\infty,z)  & (M\equiv 0 \mod 2J), \\
0 & (\mbox{others}).
\end{cases}
\end{eqnarray*}
Then, $\Zh$ is expressed in the formula
\begin{eqnarray*}
 \lim_{\tau\to i \infty}\Zh=\frac1{2^{R}N}
\sum_{\lambda,\lambdab,\nt}^{\rm even,beta}\deltam{M}{2J}
L_{\lambda\lambdab}
I_{\nu+\frac{M}{2J}\gamma_0}^{\lambda}(i\infty,z)
\bar I_{\nu+\frac{M}{2J}\gamma_0}^{\lambdab}(-i\infty,0).
\end{eqnarray*}
When we replace the $\nu+\frac{M}{2J}\gamma_0$ by $\nu$, then we can
perform the sum of $M\in \Zb_{2NJ}$.
Moreover, from the fact
\begin{eqnarray*}
 I_{m_j}^{\ell_j}(-i\infty,z)
=\deltam{m_j-\ell_j-1}{2N_j}\;y^{\frac{\ell+1}{N}-\frac12}
-\deltam{m_j+\ell_j+1}{2N_j}\;y^{-\frac{\ell+1}{N}+\frac12},
\end{eqnarray*}
it can be seen that the even condition $\ell_j+m_j\equiv 1 \mod{2}$
is included in this factor.
We obtain a formula of the $\Zh$ in this limit
\begin{eqnarray*}
 \lim_{\tau\to i \infty}\Zh=\frac1{2^{R}}
\sum_{\nu}^{\rm beta}
\prod_j \left[
\sum_{\ell_j,\bar\ell_j}
L^{(N_j)}_{\ell_j\bar\ell_j}
I_{m_j}^{\ell_j}(i\infty,z)
I_{m_j}^{\bar\ell_j}(-i\infty,0)
\right]
.
\end{eqnarray*}

So far, we treat rather general cases, but from now, 
we take an example and restrict ourselves to calculations in the example.
We consider the example which satisfies all the following conditions.
\begin{itemize}
 \item All minimal models are A type. 
So, $L_{\lambda\lambdab}=\delta_{\lambda\lambdab}$.
 \item $R=n+1$.
 \item $N_1=N_2=\dots=N_R=N$.
\end{itemize}
In other words, this example is 
the case where the associated Calabi-Yau manifold $X$ is
 the hypersurface of the form
\begin{eqnarray}
 && z_1^N+z_2^N+\dots+z_{n+1}^N=0 \mbox{ in } \Cb^{n+1}.
\label{BaseEquation}
\end{eqnarray}
We can write the finite distance condition as
$N<n+1$, which is equivalent to the condition that 
first Chern number of $X/\Cx$ is positive.
In this case, nontrivial factor of the elliptic genus
can be calculated as
\begin{eqnarray*}
&& \lim_{\tau\to i \infty}\Zh=\frac1{2^R}
\sum_{\nu}^{\rm beta}\prod_j \left[
\sum_{\ell_j}
\left(\deltam{m_j-\ell_j-1}{2N_j}\;y^{\frac{\ell+1}{N}-\frac12}
-\deltam{m_j+\ell_j+1}{2N_j}\;y^{-\frac{\ell+1}{N}+\frac12}\right)
\left(\deltam{m_j+\ell_j+1}{2N_j}-\deltam{m_j-\ell_j-1}{2N_j}\right)
\right]\\
&& \hspace{1.5cm}=
\frac{y^{-\frac{n+1}{2}}}{2^R}
\sum_{\nu}^{\rm beta}\prod_j \left[
\sum_{\ell_j}
\left(\deltam{m_j-\ell_j-1}{2N_j}\;y^{\frac{\ell+1}{N}}
+\deltam{m_j+\ell_j+1}{2N_j}\;y^{-\frac{\ell+1}{N}+1}\right)
\right].
\end{eqnarray*}
When we put $m_j=a_j+Nb_j,\ \ (b_j=0,-1;\ a_j=0,1,\dots N-1)$, then 
beta condition becomes
\begin{eqnarray*}
 \sum_j a_j\equiv 0 \mod N.
\end{eqnarray*}
We obtain the $\Zh$ in this limit
\begin{eqnarray*}
 &&\lim_{\tau\to i \infty}\Zh=\sum_{p=1}^{n}h_p y^{p-\frac{n+1}{2}},\\
 && h_p:=\sum_{
\atop{a_j=1,\dots,N-1,}{\sum_ja_j=pN}
}1=\sum_{i=0}^p(-1)^i\binom{n+1}{i}\binom{(p-i)(N-1)+p-1}{n}.
\end{eqnarray*}
We show several examples of $h_p$ for lower $n,N$ in Table \ref{TableHP}.
\begin{table}
\begin{center}
 \begin{tabular}[t]{|c||c|c|c|}\hline
 \multicolumn{4}{|c|}{$n=3$}\\\hline\hline
 $N\backslash p$ &1 &2 &3 \\ \hline\hline
 3 &0 &6 &0 \\ \hline
 4 &1 &19 &1 \\ \hline
 \end{tabular}
 \begin{tabular}[t]{|c||c|c|c|c|}\hline
 \multicolumn{5}{|c|}{$n=4$}\\\hline\hline
 $N\backslash p$ &1 &2 &3 &4 \\ \hline\hline
 3 &0 &5 &5 &0 \\ \hline
 4 &0 &30 &30 &0 \\ \hline
 5 &1 &101 &101 &1 \\ \hline
 \end{tabular}
 \begin{tabular}[t]{|c||c|c|c|c|c|}\hline
 \multicolumn{6}{|c|}{$n=5$}\\\hline\hline
 $N\backslash p$ &1 &2 &3 &4 &5 \\ \hline\hline
 3 &0 &1 &20 &1 &0 \\ \hline
 4 &0 &21 &141 &21 &0 \\ \hline
 5 &0 &120 &580 &120 &0 \\ \hline 
 6 &1 &426 &1751 &426 &1 \\ \hline
\end{tabular}
\end{center}
\caption{The values of the coefficients $h_p$ for 
$n=3,4,5,\ N=3,\dots,n+1$. 
We include the $N=n+1$ case in the table, 
despite it is suppressed by the finite distance condition.
}\label{TableHP}
\end{table}
These coefficients $h_p$
seem to coincide with the middle dimensional Hodge numbers of $X/\Cx$
except for the cohomology elements generated by cup products of a
K\"ahler form of $X/\Cx$, on which we mention below.
 
In our model, the Witten index can also be calculated as
\begin{eqnarray*}
 && \lim_{\tau\to i \infty,z\to 0}\Zh=\sum_{p=1}^{n} h_p\\
&&\hspace{2cm}=(-1)^{n+1}\left[1+\frac{(1-N)^{n+1}-1}{N}
\right]\\
&&\hspace{2cm}=(-1)^{n+1}\left[n+1+\frac{(1-N)^{n+1}-1}{N}-n\right].
\end{eqnarray*}
On the other hand, the Euler number of
the $(n-1)$-dimensional manifold $X/\Cx$ is expressed in the formula
\begin{eqnarray*}
 && \chi_{X/\Cx}=n+1+\frac{(1-N)^{n+1}-1}{N}.
\end{eqnarray*}
The Witten index of the CFT almost coincide with the Euler number
$\chi_{X/\Cx}$ of $X/\Cx$.
One of the differences of the two is the sign $(-1)^{n+1}$, but
this is not relevant. Except this difference of the sign, 
the Witten index is smaller by $n$ than the Euler number of $X/\Cx$
in our case. 
This difference may correspond to
the cohomology elements generated by cup products of the K\"ahler
form of $X/\Cx$. On $X$, these cohomology elements are absent because
they appear when we take the quotient of $X$ by $\Cx$. 
This seems the reason why the Witten index is smaller by $n$ than
the Euler number of $X/\Cx$.
\section{Conclusion and discussion}
We construct the toroidal partition function of the string theory
described by the combination of an $\Ncal=2$ Liouville theory
and multiple $\Ncal=2$ minimal models. 
This partition function is actually modular invariant, and we can conclude
that the theory exists consistently.

This string theory is thought to describe the string on
a noncompact singular Calabi-Yau manifold.
To check this proposition,
we also calculate the elliptic genus of this theory
and the Witten index.

The Euler number defined from
non-trivial factor of the Witten index in the CFT 
seems to be that of the non-vanishing 
elements of the cohomology.
In the case of a singular manifold, there are
vanishing elements of the cohomology, which are supported on
the singular point and reflect the structure of singularities.

The fact that
the vanishing elements of the cohomology cannot be seen,
is probably related
to our method of construction in which we include
only the states in ``principal continuous series'' of the
$SL(2)$ theory.
If we can include some ``discrete series''
(but it is difficult\cite{KS0001})
, the structure of the singularities
might be seen in the CFT.

Another reason is 
that we treat the $\Ncal=2$ Liouville theory as free field theory in
this paper.
It is mentioned in \cite{ES0009} that if we treat appropriately
the effect of Liouville potential, the Witten index does not
vanish and gives the Euler number including the vanishing elements of
the cohomology.
In this paper, since we treat the case of $\mu=0$ and not deformed
singularity, the vanishing elements of the cohomology actually 
vanish and it is consistent with the vanishing Witten index.

We may not be able to use our result to 
analyze the structure of the singularities,
but we can use it to analyze the string on the 
positively curved manifold $X/\Cx$. Especially it is
interesting to analyze the D-branes wrapped on infinite cycle
in this noncompact Calabi-Yau manifold through the recipes of
 boundary states in the CFT \cite{EGKRS0005,Ler0006,LLS0006,ES0009}
 as the case of the ordinary Gepner models
\cite{RS9712,BDLR9906,NN0001,Sug0003A}.

\subsection*{Acknowledgement}

I would like to thank Tsuneo Uematsu and Katsuyuki Sugiyama
for useful discussions and encouragements.
I would also like to thank to the organizers 
of Summer Institute 2000 at Yamanashi,
Japan, 7-21 August, 2000, and the participants of it, especially,
Michihiro Naka, Masatoshi Nozaki, Yuji Sato and Yuji Sugawara
for useful discussions.

This work is supported in part by JSPS Research Fellowships for
Young Scientists.

\newpage
\section*{Appendix A. \ \ \ Theta functions and characters}
We use the following notations in this  paper.
\begin{eqnarray*}
 &&\e{x}:=\exp(2\pi i x),\\
 &&\deltam{m}{N}:=
\begin{cases}
 1 & (m\equiv 0 \mod N),\\
 0 & ({\rm others}),
\end{cases}
\end{eqnarray*}
where $m$ and $N$ are integers.
The useful formula is
\begin{eqnarray*}
 &&\sum_{j\in \Zb_N}\e{\frac{jm}{N}}=N\deltam{m}{N},
\end{eqnarray*}
where $m$ and $N$ are integers.

The SU(2) classical theta functions are defined as
\begin{eqnarray*}
&& \Th_{m,k}(\tau,z)=\sum_{n\in \Zb}q^{k\left(n+\frac{m}{2k}\right)^2}
 y^{k\left(n+\frac{m}{2k}\right)},\\
\end{eqnarray*}
where $q:=\e{\tau},y:=\e{z}$.
The Jacobi's theta functions are also defined as
\begin{eqnarray*}
&& \theta_{1}(\tau,z):=i\sum_{n\in \Zb}(-1)^n q^{\left(n-\frac{1}{2}\right)^2}
 y^{\left(n-\frac{1}{2}\right)},
 \theta_{2}(\tau,z):=\sum_{n\in \Zb} q^{\left(n-\frac{1}{2}\right)^2}
 y^{\left(n-\frac{1}{2}\right)},\\
&& \theta_{3}(\tau,z):=\sum_{n\in \Zb} q^{n^2}y^{n},\hspace{2.7cm}
 \theta_{4}(\tau,z):=\sum_{n\in \Zb}(-1)^n q^{n^2}y^{n}.\\
\end{eqnarray*}
Two kinds of theta functions are related by equations
\begin{eqnarray*}
 && 2\Th_{0,2}=\theta_3+\theta_4 ,\ 2\Th_{1,2}=\theta_2+i\theta_1, \\
 && 2\Th_{2,2}=\theta_3-\theta_4 ,\ 2\Th_{3,2}=\theta_2-i\theta_1.
\end{eqnarray*}

The Dedekind $\eta$ function is represented as an infinite product
\begin{eqnarray*}
 \eta(\tau):=q^{\frac1{24}}\prod_{n=1}^{\infty}(1-q^n).
\end{eqnarray*}

The character $\chi_s(\tau,z),\ s=0,1,2,3$ 
of $\widehat{SO(d)}_1$ for $d/2 \in 2\Zb+1$ can be
expressed as
\begin{eqnarray*}
 && \chi_0(\tau,z)=\frac{\theta_3(\tau,z)^{d/2}
+\theta_3(\tau,z)^{d/2}}{2\eta(\tau)^{d/2}},\\
 && \chi_1(\tau,z)=\frac{\theta_2(\tau,z)^{d/2}
+(i\theta_1(\tau,z))^{d/2}}{2\eta(\tau)^{d/2}},\\
 && \chi_2(\tau,z)=\frac{\theta_3(\tau,z)^{d/2}
-\theta_3(\tau,z)^{d/2}}{2\eta(\tau)^{d/2}},\\
 && \chi_3(\tau,z)=\frac{\theta_2(\tau,z)^{d/2}
-(i\theta_1(\tau,z))^{d/2}}{2\eta(\tau)^{d/2}}.
\end{eqnarray*}

Let us denote the character of 
a Verma module $(\ell,m,s)$ in the level $(N-2)$ minimal model 
as $\chi_m^{\ell,s}(\tau,z)$. This character satisfies equivalence
relations
\begin{eqnarray*}
 \chi_m^{\ell,s}=\chi_{m+2N}^{\ell,s}=\chi_m^{\ell,s+4}
=\chi_{m+N}^{N-2-\ell,s+2}.
\end{eqnarray*}
The explicit form of this character is written in \cite{Gep88}.

We collect the modular properties of these functions.
Under the T transformations, they behave as
\begin{eqnarray*}
 && \Th_{m,k}(\tau+1,z)=\e{\frac{m^2}{4k}}\Th_{m,k}(\tau,z),\\
 && \theta_1(\tau+1,z)=\e{\tfrac18}\theta_1(\tau,z), 
 \qquad \theta_2(\tau+1,z)=\e{\tfrac18}\theta_2(\tau,z),\\
 && \theta_3(\tau+1,z)=\theta_4(\tau,z), 
 \qquad \theta_4(\tau+1,z)=\theta_3(\tau,z),\\
 && \eta(\tau+1)=\e{1/24}\eta(\tau),\\
 && \chi_s(\tau+1,z)=\e{\frac{s^2}{8}-\frac{d}{48}}\chi_s(\tau,z),\\
 && \chi_m^{\ell,s}(\tau+1,z)=\e{\frac{\ell(\ell+2)}{4N}-\frac{m^2}{4N}
+\frac{s^2}{8}-\frac{N-2}{8N}}\chi_m^{\ell,s}(\tau,z),
\end{eqnarray*}
and for S transformations, they have modular properties
\begin{eqnarray*}
 &&\Th_{m,k}(-1/\tau,z/\tau)=
\sqrt{-i\tau}\e{\frac{k}{4}\frac{z^2}{\tau}}
\sum_{m'\in \Zb_{2k}}\frac{1}{\sqrt{2k}}\e{-\frac{mm'}{2k}}
\Th_{m',k}(\tau,z),\\
 &&\theta_1(-1/\tau,z/\tau)=-i\sqrt{-i\tau}\e{\frac12 \frac{z^2}{\tau}}
\theta_1(\tau,z),
\qquad \theta_2(-1/\tau,z/\tau)=\sqrt{-i\tau}\e{\frac12 \frac{z^2}{\tau}}
\theta_4(\tau,z),\\
 &&\theta_3(-1/\tau,z/\tau)=\sqrt{-i\tau}\e{\frac12 \frac{z^2}{\tau}}
\theta_3(\tau,z),
\qquad \theta_4(-1/\tau,z/\tau)=\sqrt{-i\tau}\e{\frac12 \frac{z^2}{\tau}}
\theta_2(\tau,z),\\
&& \eta(-1/\tau)=\sqrt{-i\tau}\eta(\tau),\\
&&\chi_s(-1/\tau,z/\tau)=\e{\frac{d}{4}\frac{z^2}{\tau}}
\sum_{s'=0}^{3}\frac 12 \e{-\frac d2 \frac{ss'}{4}}\chi_{s'}(\tau,z),\\
 && \chi_m^{\ell,s}(-1/\tau,z/\tau)=
\e{\frac{N-2}{2N}\frac{z^2}{\tau}}
\frac{1}{\sqrt{8N}} \sum_{\ell,m,s}^{\rm even}
A_{\ell\ell'}\e{-\frac{ss'}{4}+\frac{mm'}{2N}}
\chi_{m'}^{\ell',s'}(\tau,z),\\
&&A_{\ell\ell'}=\sqrt{\frac2N}\sin\left[\pi\frac{(\ell+1)(\ell'+1)}{N}\right],
\end{eqnarray*}
where the sum $\sum_{\ell,m,s}^{\rm even}$ means that 
 $\ell+m+s \equiv 0\mod 2$ for $(\ell,m,s)$.

We use the notation $f(\tau)$ for a function $f(\tau,z)$
of $\tau,z$ with substituting $z=0$
\begin{eqnarray*}
 f(\tau):=f(\tau,z=0).
\end{eqnarray*}

\newpage
\providecommand{\href}[2]{#2}\begingroup\raggedright\endgroup

\end{document}